# Planck-Einstein-de Broglie type relations for the acoustic waves

Ion Simaciu[1,a], Zoltan Borsos[1,b], Gheorghe Dumitrescu[2], Anca Baciu[1]

[1] Petroleum-Gas University of Ploieşti, Ploieşti 100680, Romania

[2] High School Toma N. Socolescu, Ploieşti, Romania

## Abstract

*In this paper we prove, by expressing the energy as a function of the wave propagation speed, it is highlighted the existence of an equivalent mass of the wave, as well as of an Einstein type relations between the energy and this mass. Also, we establish a relation between angular frequency and energy similar to that of the Planck relation. For the propagating wave, there is a de Broglie type relationship between the linear momentum and the action variable (the angular momentum), i.e. the wave linear momentum is proportional to the wave number, the proportionality coefficient being the action.*

## 1. Introduction

In earlier papers [1, 2, 3] authors emphasized a similitude between our world, i.e. the electromagnetic world and the acoustic world. In our world, the fundamental interactions i.e. the information and the energy exchange, is supposed to travel with a finite speed. This is the speed of light in vacuum.

In the acoustic world, which pertain to an elastic medium, the maximum speed is the one of the mechanical waves. In this latter world, the observers, and also their devices, interact through the mechanical waves. The picture of this world is formed by the waves which travel the medium, the standing waves, the wave-packets generated by diffraction-interference, the flow of the fluid, the bubbles generated in medium, and so on.

The main reason behind this concept being introduced in physics was to attempt to describe the processes which accompany the collapsing of mass, i.e. the black holes, and the emission of the black holes. An observer from this acoustic world can infer a Lorentz type invariance of the parameters of the processes which take place and also the existence of an acoustic black hole, i.e. dumb hole [4, 5, 6].

The aim of this paper is to make easier to understand wave-particle duality concept. Also this physico-mathematical approach to classical of phenomena, i.e. acoustic wave propagation in a finite medium, will enable us to develop an analogy between acoustic world and electromagnetic world for the understanding of phenomena dumb hole and black hole. Studying the processes taking place in the acoustic world leads to duality wave-particle concept. We will establish some relations between the parameters of the acoustic disturbances (waves and wave packets) and the ones of the corpuscular systems. We will provide a relation similar to that of mass-energy equivalence of Einstein, a relation between angular frequency and energy similar to that of the Planck relation and a relation of the de Broglie type between the wavelength and the linear momentum. All these parameters, corpuscular and undulatory, are correlated through the speed of the mechanical wave and the angular momentum. Our work is done based on the background of classical mechanics using the theory of mechanical oscillations and mechanical waves, as well as analytical mechanics.

[a] sion@upg-ploiesti.ro; isimaciu@yahoo.com

[b] borsos.zoltan@gmail.com

In the second section we establish the link between the angular frequency of the allowed modes $\omega_j$ of a bar embedded at both ends and the natural angular frequency of bar $\omega_b$. The portion of the medium with the size, in the direction of propagation, equal to one half wavelength acts as a macroscopic oscillator. The wave is an assembly of macroscopic oscillators having a half wavelength size. Because of the microscopic point of view, the bar is a system of microscopic/atomic oscillators, we establish the link between the angular frequency for atomic oscillators $\omega_a$, the angular frequency of the allowed modes $\omega_j$ and the natural angular frequency of bar $\omega_b$.

In the third section, we calculate the energy of the wave that propagates and of the stationary wave, by calculating the parameters of the portion of the medium with the size, in the direction of propagation, equal to one half wavelength. For these oscillator, one can define and calculate the action variable (i.e. the canonical action variable) corresponding to the oscillation energy. The same parameters are calculated based on the energy and the action formulae of the constituent particles of the medium (in particular, atoms and molecules). We will provide a relation between angular frequency and energy similar to that of the Planck relation [7, Ch. 1.1; 8, Ch. 41.3].

In the fourth section, we express the energy as a function of the speed of the propagating wave. It highlights the existence of an equivalent mass of the wave and of Einstein type dependances between the energy and this mass [8, Ch. 15.9]. One introduces the linear momentum of the wave (for the wave that propagates). Between this linear momentum and the generalized one (the action variable) there is a de Broglie type relation (i.e. the linear momentum of the wave is proportional to the wave number, the coefficient of proportionality being the action) [7, Ch. 1.3; 14, Ch.4.].

The section 5 includes the conclusions of the paper.

## 2. The mechanical wave in a finite physical system

In this section we study the standing waves in a finite mechanical system. The phenomenon is analyzed both at macroscopic level and the microscopic level.

### 2.1. Stading Waves in a bar fixed at both ends

Be there a bar with the section $S = l_x l_y$ and length $l_z \gg l_x, l_y$, mass density $\rho$ and elasticity modulus (Young' modulus) $Y$. Assuming the law of Hooke [9, Ch. 38.1], the elastic constant of the bar $K$ depends on the elasticity modulus $Y$, according to the relation

$$K = \frac{Y l_x l_y}{l_z}. \tag{1}$$

The longitudinal waves propagate in the bar with the speed [9, Ch. 38.3]

$$u = \sqrt{\frac{Y}{\rho}}, \tag{2a}$$

(if one neglects the transversal deformation) or, when replacing the Young modulus given by (1),

$$u = \sqrt{\frac{K l_z}{l_x l_y \rho}}. \tag{2b}$$

If the bar is fixed at both ends, then standing waves are formed within it [10, Ch. 2]. The allowed modes are given by the expression of the wave number (that is the length of bar is equal to an integer number of half-wavelengths) $l_z = j \lambda_j / 2$

$$k_j = j \frac{\pi}{l_z}; \; j = 1, 2, 3, \ldots, \tag{3}$$

From the definition of the phase speed for a wave

$$u = \frac{\omega}{k}. \tag{4}$$

and the relation (3) of the modes, there results the expresion for the angular frequency of the allowed modes

$$\omega_j = u k_j = j \frac{\pi}{l_z} \sqrt{\frac{K l_z}{l_x l_y \rho}}, \tag{5a}$$

or, with the bar mass $m = l_x l_y l_z \rho$,

$$\omega_j = u k_j = j \pi \sqrt{\frac{K}{m}}. \tag{5b}$$

If we consider the measure

$$\omega_b (l_z) = \sqrt{\frac{K}{m}}, \tag{6}$$

to be the angular frequency intrinsic to a mechanical oscillator with a $K$ elastic constant and the mass concentrated in a single point, there results, with (5b)

$$\omega_j = u k_j = j \pi \omega_b, \tag{7}$$

The angular frequency given by relation (6) is the minimum angular frequency of the bar; it serves as a guide for the mechanical waves.

## 2.2. Macroscopic oscillator

As for the standing waves and the waves that propagate, we can separate the portions of the medium in which propagate having a length equal to one half wavelenght or a quarter wavelenght (the case when the bar is freely at one end).

The portion of the medium with this size, in the direction of propagation, acts as a macroscopic oscillator.

The specific angular frequency of these microscopic oscillators is close in size to the angular frequency wave.

We calculate the angular frequency of a bar with the section $S = l_x l_y$ and length $\lambda/2$. The elastic constant corresponding to this bar will be, according to the relation

$$K_{\lambda/2} = \frac{2Y l_x l_y}{\lambda}. \quad (8)$$

When replacing in relation (6) the expression of the elastic constant (8), there results the specific angular frequency of the this bar

$$\omega_{\lambda/2} = \omega_b(\lambda/2) = \sqrt{\frac{K_{\lambda/2}}{m_{\lambda/2}}} = \frac{\omega}{\pi}. \quad (9)$$

By the same method, the pulsation corresponding to a section of length $\lambda/4$ is

$$\omega_{\lambda/4} = \omega(\lambda/4) = \frac{2\omega}{\pi}. \quad (10)$$

This specific angular frequencies are not dependent on section $S = l_x l_y$.

## 2.3. Microscopic modelling of the oscilations

From a microscopic point of view, the bar is a system of particles with the masa $m_a$ arranged in a cubic crystal lattice with the constant $a$ and connected by elastic forces. From a mechanical point of view, one can shape the bar as a three-dimensional system of particles connected by elastic springs with the constant $K_a$.

The particles from the transverse plane $Oxy$ with the mass

$$m_{xy} = \frac{S_{xy}}{a^2} m_a = \frac{l_x l_y}{a^2} m_a = N_{xy} m_a, \quad (11)$$

are connected to other particles from another transverse plane, but moved with the lattice constant $a$ by means of $N_{xy}$ parallel springs. One can reduce the three-dimensional system to a monodimensional system of particles with the mass $m_{xy}$ linked by equivalent springs having the constant

$$K_z = K_{parallel} = N_{xy} K_a. \quad (12)$$

There follows that the bar was reduced by means of modelling to a system of

$$N_z = \frac{l_z}{a} \quad (13)$$

bodies with the mass $m_{xy}$ connected by elastic springs with the constant $K_z$.

According to the oscillations theory for this system, the dispersion relation of the specific oscillation modes [10, p. 80] is

$$\omega(k_j) = 2\sqrt{\frac{K_z}{m_{xy}}} \sin\frac{ak_j}{2}, \qquad (14)$$

with the wave vectors given by the relation (3) and

$$j = 1, 2, \ldots, N_z. \qquad (15)$$

When replacing (11) and (12) in (14), there results

$$\omega(k_j) = 2\sqrt{\frac{K_a}{m_a}} \sin\frac{ak_j}{2} = 2\omega_a \sin\frac{ak_j}{2}, \qquad (16)$$

with the natural angular frequency for atomic oscilator

$$\omega_a = \sqrt{\frac{K_a}{m_a}}. \qquad (17)$$

When approximating the waves with a wavelength much higher than $a$, there results

$$\omega(k_j) = ak_j\sqrt{\frac{K_a}{m_a}} = k_j\sqrt{\frac{aK_a}{(m_a/a)}}, \qquad (18a)$$

or

$$\frac{\omega(k_j)}{k_j} = \sqrt{\frac{aK_a}{(m_a/a)}} = u, \qquad (18b)$$

which leads back to relation (4).

If one writes the relation (18a) with the formula

$$\omega_j = j\frac{\pi a}{l_z}\sqrt{\frac{K_a}{m_a}} = j\pi\sqrt{\frac{a^2 K_a}{m_a l_z^2}} \qquad (19)$$

and considers the fact that for a monodimensional series of springs there exists relation [10, p. 81]

$$Kl_z = K_z a = (N_{xy} K_a)a, \qquad (20)$$

there follows

$$\omega_j = j\pi\sqrt{\frac{al_z K}{N_{xy} m_a l_z^2}} = j\pi\omega_b, \qquad (21)$$

which is, in fact, relation (7).

The dispersion relation for high wavelengths is

$$\omega_j = j\frac{\omega_a}{(l_z/a)} = j\frac{\omega_a}{N_z} \qquad (22)$$

and, considering expression (21), the relation

$$\omega_b = \frac{\omega_a}{\pi N_z}. \qquad (23)$$

connects the natural angular frequency of bar to the natural angular frequency of atomic oscillator.

## 3. The oscillation energy of the bar

### 3.1. Standing wave energy of the bar

For a standing longitudinal wave $\Psi_s(z,t) = q_{sz} = q_{0sz}\sin(kz)\sin(\omega t)$ [11], each element with the length $dx$ and mass $dm = \rho l_x l_y dz$ performs an oscillatory motion and has the kinetic energy

$$dE_k = \frac{1}{2}dm\left(\frac{\partial q_{sz}}{\partial t}\right)^2 = \frac{1}{2}dm q_{0sz}^2 \omega^2 \sin^2(kz)\cos^2(\omega t) \qquad (24a)$$

and the potential energy

$$dE_p = \frac{1}{2}dm u^2\left(\frac{\partial q_{sz}}{\partial z}\right)^2 = \frac{1}{2}dm q_{0sz}^2 \omega^2 \cos^2(kz)\sin^2(\omega t). \qquad (24b)$$

Therefore, total energy is

$$dE = dE_k + dE_p = \frac{1}{2}dm q_{0sz}^2 \omega^2 \left[\sin^2(kz)\cos^2(\omega t) + \cos^2(kz)\sin^2(\omega t)\right], \qquad (25)$$

With (25), the oscillation energy of the bar is

$$E = \frac{1}{2}\rho l_x l_y q_{0sz}^2 \omega^2 \int_0^{l_z}\left[\sin^2(kz)\cos^2(\omega t) + \cos^2(kz)\sin^2(\omega t)\right]dz = \\ \frac{1}{4}\rho l_x l_y l_z q_{0sz}^2 \omega^2 = \frac{1}{4}m q_{0sz}^2 \omega^2. \qquad (26)$$

For the mode $j$, with $\lambda_j = 2l_z/j$, the energy of the bar that oscillates stationarily is

$$E_{sj} = \frac{j}{8}\rho l_x l_y q_{0szj}^2 \omega_j^2 \lambda_j = j\left(\frac{\lambda_j}{2}l_x l_y n\right)\left(\frac{1}{4}m_a q_{0szj}^2 \omega_j^2\right) = jN_{\lambda_j/2}E_{asj} \qquad (27)$$

with $jN_{\lambda_j/2}$ being the number of microscopic/ atomic oscillators in the bar and

$$E_{asj} = \frac{1}{4}m_a q_{0szj}^2 \omega_j^2 \qquad (28)$$

the average energy of a microscopic/atomic oscillator.

As the wavelength can be expressed, based on the propagation speed of the waves $u$ and angular frequency $\omega_j$

$$\lambda_j = \frac{2\pi u}{\omega_j}, \qquad (29)$$

there follows that, when replacing in (27), the energy of the bar, as a macroscopic oscillator, is proportional with the angular frequency

$$E_{sj} = j\left(\frac{\pi}{4}\rho l_x l_y q_{0szj}^2 u\right)\omega_j = jJ_{s\lambda_j/2}\omega_j. \qquad (30)$$

In analytical mechanics [12, p. 431], the energy of the oscillator in harmonic coordinates has the expression (30), where

$$J_{s\lambda_j/2} = \left(\frac{\lambda_j}{2}l_x l_y n\right)\left(\frac{1}{4}m_a q_{0szj}^2 \omega_j\right) = N_{\lambda_j/2}J_{saj} \qquad (31)$$

is the action variable and

$$E_{s\lambda_j/2} = \left(\frac{\pi}{4}\rho l_x l_y q_{0szj}^2 u\right)\omega_j = J_{s\lambda_j/2}\omega_j \qquad (32)$$

is the energy of the macroscopic oscillator corresponding to a half wavelength.

In the expresion of relation (31)

$$J_{saj} = \frac{1}{4}m_a q_{0szj}^2 \omega_j \qquad (33)$$

is *the action variable* (the canonical action variable) of the microscopic/atomic oscillator with the energy calculated with the relation (26).

For a given material, the energy and the action variable of the microscopic oscillator have the lowest value corresponding to thermal oscillations, for solids, and to thermal motion, in fluids with the temperature $T \neq 0\text{K}$.

For such a medium, the disturbances (waves and wave packets formed by means of interference-diffraction) form an "acoustic world". The disturbances are correlated by the maximum speed which is the speed of the waves in the medium $u$. The fluctuations produced by waves in the material overlap these thermal oscillations and movements. In this world, the group speed of perturbations (i.e. the speed with which the amplitudes propagate) is less than the speed of the waves $\upsilon = \upsilon_g \leq u$. There follows that the lowest values of the action variable $E_{\min}$ and of the $J_{\min}$

$$J_{\min} = J_{aT} = \frac{1}{2}m_a q_{0zT}^2 \omega_T, \qquad (34)$$

$$E_{\min} = E_{aT} = \frac{1}{2}m_a q_{0zT}^2 \omega_T^2 = J_{aT}\omega_T \qquad (35)$$

are the ones corresponding to thermal oscillations of the microscopic oscillators. The limit value for the amplitude is the average length between atoms, in the fluid, or the lattice constant, in the solid, $q_{op} \leq a$. Also, the limit value for the speed of oscillation is $q_{op}\omega_T \leq u$.

With these values the action variable and the energy becomes $J_{\min} = J_{aT} \cong m_a a u$, $E_{\min} = E_{aT} \cong m_a u^2$. It follows a classical interpretation of the Planck constant [13] knowing that also for the mechanical oscillations $E = J_{aT}\omega_T = \hbar \omega_T$ [14, Ch. 4].

### 3.2. The oscillation energy of the bar during the wave propagation

For an wave with the angular frequency $\omega$ and amplitude $q_{0z} \neq q_{0sz}$, that propagates in a bar (where the length of the bar is $l_z \gg \lambda$), the bar can be considered as a system of oscillators with the length $\lambda/2$. The bar vibrates according to $\Psi(z,t) = q_z = q_{0z}\sin(\omega t - kz)$. When using the same relations, namely (24) and (25), for the kinetic energy, potential and total, the corresponding energy of a half wavelength is

$$E_{\lambda/2} = \left(\frac{\pi}{2}\rho l_x l_y q_{0z}^2 u\right)\omega = J_{\lambda/2}\omega \tag{36}$$

and the action variable is

$$J_{\lambda/2} = \frac{\pi}{2}\rho l_x l_y q_{0z}^2 u . \tag{37a}$$

or

$$2\pi J_{\lambda/2} = \pi^2 \rho l_x l_y q_{0z}^2 u . \tag{37b}$$

The action variable of the macroscopic oscillator having the length equal to one half wavelenght is independent of the wavelength. The relations (32) and (36) for macroscopic oscillator is similar to the Planck's relation [7, Ch. 1.1; 8, Ch. 41.3].

The energy and the canonical action variable can be also expressed based on microscopic measures, namely the mass of the atoms $m_a$ and their volume number density $n$

$$E_{\lambda/2} = \left(\frac{\pi}{2}nm_a l_x l_y q_{0z}^2 u\right)\omega = \left(nl_x l_y \frac{\lambda}{2}\right)\left(\frac{\pi}{\lambda}m_a q_{0z}^2 u\omega\right) = N_{\lambda/2} E_a \tag{38}$$

$$J_{\lambda/2} = \left(nl_x l_y \frac{\lambda}{2}\right)\left(\frac{\pi}{\lambda}m_a q_{0z}^2 u\right) = N_{\lambda/2} J_a , \tag{39}$$

with $N_{\lambda/2}$ the number of atoms from the volume $V = l_x l_y \lambda/2$, the energy $E_a = (m_a q_{0z}^2 \omega^2)/2 = J_a \omega$ and the action variable $J_a = m_a q_{0z}^2 \omega/2 = \pi m_a q_{0z}^2 u/\lambda$ corresponding to the microscopic/atomic oscillator.

## 4. The oscillation energy of the medium and Einstein–de Broglie type relations

### 4.1. Einstein–de Broglie type relations for the energy and the equivalent mass of the wave

The energy of the macroscopic oscillator resulting from relation (36) can be expressed based on the speed of the wave, if we consider relation (4)

$$E_{\lambda/2} = \left(\frac{\pi}{2}\rho l_x l_y k q_{0z}^2\right)u^2 = \left(\pi^2 \rho l_x l_y \frac{q_{0z}^2}{\lambda}\right)u^2 . \tag{40}$$

By introducing the concept of *equivalent mass* of the macroscopic oscillator corresponding to a half wavelength

$$m_{u\lambda/2} = \frac{\pi}{2}\rho l_x l_y k q_{0z}^2 = \left(\rho l_x l_y \frac{\lambda}{4}\right)\left(\frac{2\pi q_{0z}}{\lambda}\right)^2 = \frac{m_{\lambda/2}}{2}\left(\frac{2\pi q_{0z}}{\lambda}\right)^2, \qquad (41)$$

with the mass $m_{\lambda/2} = \rho l_x l_y \lambda / 2$ for the substance of the macroscopic oscillator, we can write an Einstein [8, Ch. 15.9] type expression of the

$$E_{\lambda/2} = m_{u\lambda/2} u^2. \qquad (42)$$

The linear momentum on $Oz$ direction corresponding to the macroscopic oscillator is

$$p_{z\lambda/2} = m_{u\lambda/2} u. \qquad (43)$$

For the wave, there can be also defined an internal linear momentum, as the average square oscillation momentum (temporal and spacial averaging) of the atomic oscillators (generally, microscopic oscillators). The momentum of an atomic oscillator in a wave is

$$p_a = m_a \frac{\partial q_z}{\partial t} = m_a q_{0z} \omega \cos(\omega t - kz), \qquad (44)$$

The average square oscillation momentum is

$$P_a = \sqrt{\langle p_a^2 \rangle_{z,t}} = \frac{1}{2} m_a q_{0z} \omega. \qquad (45)$$

The internal macroscopic momentum is the total momentum corresponding to the oscillators from a half wavelength

$$P_{\lambda/2} = \left(n l_x l_y \frac{\lambda}{2}\right) P_a = \left(\frac{\lambda}{2\pi q_{0z}}\right) p_{z\lambda/2} \qquad (46a)$$

or

$$P_{\lambda/2} = \frac{1}{2}\left(\frac{2\pi q_{0z}}{\lambda}\right) m_{\lambda/2} u \qquad (46b)$$

and is much higher than the wave linear momentum (43), for $\lambda \gg q_{0z}$.

The ratio between the linear momentum of the oscillator (43) and the action variable (37), is

$$\frac{p_{z\lambda/2}}{J_{\lambda/2}} = k, \qquad (47)$$

namely a *de Broglie type relation between the momentum and the wave vector* analogous to that for the photon [7, Ch. 1.3] and the phonon [14, Ch. 4].

### 4.2. Einstein type relations and the equivalent mass of the stationary wave

For the harmonic $j$ of the standing waves in the bar, the macroscopic oscillator corresponding to the bar is a set of $j$ macroscopic oscillators corresponding to a half wavelength, which, according to relations (30) and (31), has the energy

$$E_{s\lambda_j/2} = \left(\frac{\pi^2}{2}\rho l_x l_y \frac{q_{0szj}^2}{\lambda_j}\right) u^2 .\qquad(48)$$

The equivalent mass corresponding to this oscillator, is

$$m_{us\lambda_j/2} = \frac{m_{\lambda_j/2}}{4}\left(\frac{2\pi q_{0szj}}{\lambda_j}\right)^2 = m_{\lambda_j/2}\left(\frac{2\pi q_{0zj}}{\lambda_j}\right)^2 = 2m_{u\lambda_j/2} .\qquad(49)$$

The linear momentum on $Oz$ direction and the total average oscillation linear momentum are zero.

For the standing wave, we can define the internal linear momentum as the average square oscillation momentum (temporal and spacial averaging) of the atomic oscillators (generally, microscopic oscillators). The momentum for an atomic oscillator in the stationary wave is

$$p_{saj} = m_a \frac{\partial q_{szj}}{\partial t} = m_a q_{0szj} \omega_j \sin(k_j z)\sin(\omega_j t) .\qquad(50)$$

The square atomic momentum, averaged in time and space, is

$$P_{saj} = \sqrt{\langle p_{aj}^2\rangle_{z,t}} = \frac{1}{2}m_a q_{0szj}\omega_j .\qquad(51)$$

The internal macroscopic momentum is the total momentum corresponding to the oscillators from a half wavelength

$$P_{s\lambda_j/2} = \left(n\frac{\lambda_j}{2}l_x l_y\right)P_{saj} = 2\left(\frac{\lambda_j}{2\pi q_{0szj}}\right)(m_{us\lambda_j/2}u) = \left(\frac{2\pi q_{0szj}}{\lambda_j}\right)\frac{m_{\lambda_j/2}u}{2} .\qquad(52)$$

The expression for this momentum differs from expressions (46) in the values of amplitude ($q_{0szj} \neq q_{0zj}$ for the standing wave with the wavelength $\lambda_j$ - the amplitudes are connected through the relation $q_{0szj} = 2q_{0zj}$) of the wavelength $\lambda_j \neq \lambda$.

## 5. Conclusions

The expression of the energy carried by a mechanical wave depending on the wave propagation velocity shows the existence of an equivalent mass of the wave. The two quantities are connected through an Einstein type relation. This mass is the analogue of the relativistic mass of the electromagnetic wave. For this kind of wave exist and a relation between angular frequency and energy similar to that of the Planck relation. The mechanical wave has also a linear momentum and an action which are connected through a de Broglie type relation analogous to the one between the momentum and the wavelength corresponding to the photon and phonon.

The standing wave is characterized by an energy and a rest equivalent mass connected through an Einstein type relation. There are also a relation between angular frequency and energy similar to that of the Planck relation.

In further papers we will study the relations of Planck-Einstein-de Broglie type for the wave packets and some consequences of them.